\documentclass[doublespacing]{elsart} 

\usepackage{graphicx} 

\usepackage{amssymb} 

\def\be{\begin{equation}}  
\def\ee{\end{equation}}  
\def\ba{\begin{eqnarray}}  
\def\ea{\end{eqnarray}}  
\def\bc{\begin{center}}  
\def\ec{\end{center}}  
\def\sgn{{\rm sgn}}

\begin{document} 

\begin{frontmatter} 


\title{Non-linear graphene optics 
for terahertz applications}
\author{S. A. Mikhailov}
\ead{sergey.mikhailov@physik.uni-augsburg.de} 
\address{Institute of Physics, University of Augsburg, D-86135 Augsburg, Germany} 





\begin{abstract} 
The linear electrodynamic properties of graphene -- the frequency-dependent conductivity, the transmission spectra and collective excitations -- are briefly outlined. The non-linear frequency multiplication effects in graphene are studied, taking into account the influence of the self-consistent-field effects and of the magnetic field. The predicted phenomena can be used for creation of new devices for microwave and terahertz optics and electronics.  
\end{abstract} 

\end{frontmatter} 

\section{Introduction} 
\label{intro} 

\begin{figure}  
\includegraphics[width=6cm]{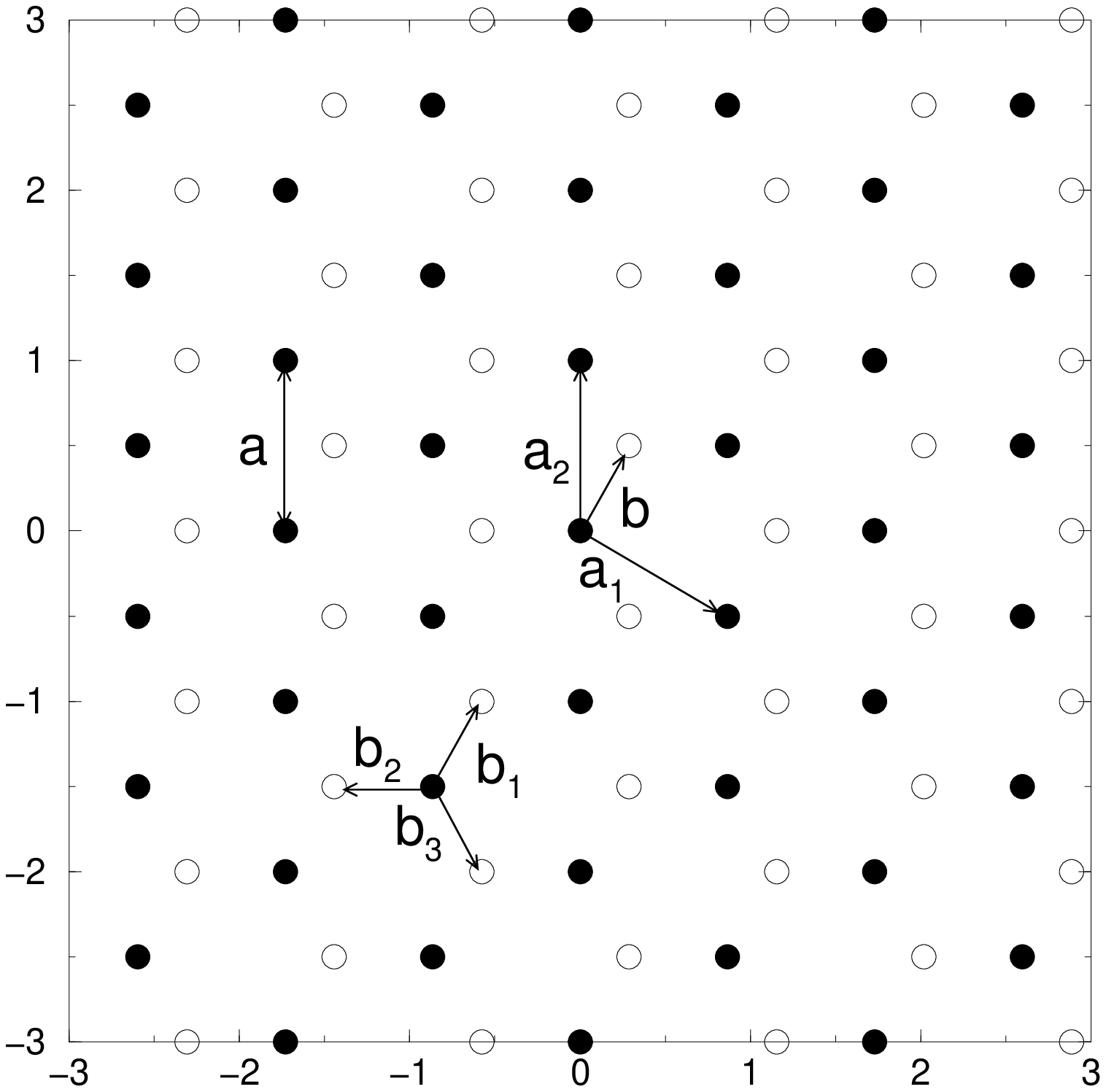}\hfill\includegraphics[width=7.5cm]{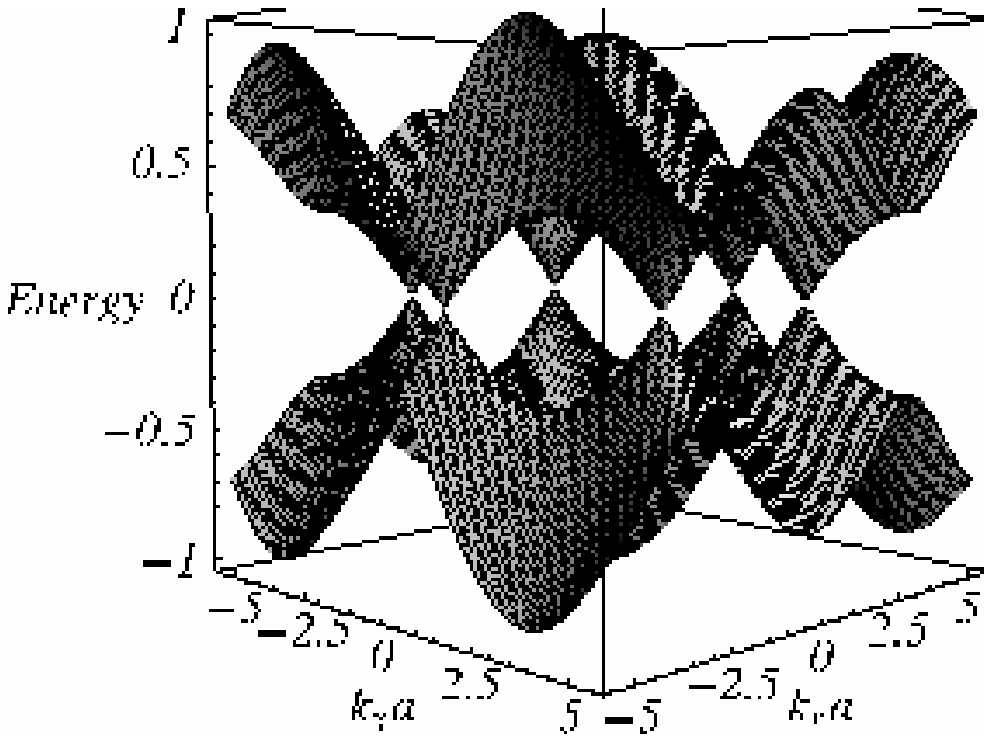}\\  
\caption{\label{latt} {\bf Left:} The honeycomb lattice of graphene. The basis lattice vectors are ${\bf a}_1=(\sqrt{3}/2,-1/2)a$ and ${\bf a}_2=(0,1)a$; the ${\bf b}$ vectors are ${\bf b}_j=(a/2)(\cos\phi_j,\sin\phi_j)$, $\phi_j=\pi/ 3+j2\pi/ 3$, $j=0,1,2$, where $a=2.46$\AA \  is the lattice constant.  {\bf Right:} The band structure of graphene electrons (\ref{sp}). The energy is normalized to the full width of the energy band $\Delta=2\sqrt{3}\hbar V/a$. The six Dirac points lie at ${\bf Q}_j=(4\pi/3a)(\cos\phi_j,\sin\phi_j)$, $\phi_j=\pi/ 6+j\pi/ 3$, $j=0,\dots,5$.
}  
\end{figure}

Graphene is a new material experimentally discovered about three years ago \cite{Novoselov05,Zhang05}. This is a monolayer of carbon atoms packed in a dense two-dimensional (2D) honeycomb lattice, Figure \ref{latt}, left panel. The spectrum of electrons in graphene  
\be  
E_{{\bf k}l}=(-1)^l \Delta\left|e^{i{\bf k}{\bf b}_1}  
+e^{i{\bf k}{\bf b}_2}  
+e^{i{\bf k}{\bf b}_3}\right|/3, \ l=1,2, \label{sp} 
\ee  
can be calculated in the tight-binding approximation taking into account the symmetry of the lattice (here ${\bf k}=(k_x,k_y)$ is the electron wavevector; the vectors ${\bf b}_i$ are defined in the caption of Figure \ref{latt}). It consists of two bands ($l=1$ and 2) which touch each other at the six corners of the hexagon shaped Brillouin zone, Figure \ref{latt}, right panel. In uniform and undoped graphene at zero temperature, the lower band $E_{{\bf k}1}$ is fully occupied while the upper band $E_{{\bf k}2}$ is empty, and the Fermi level goes through these six, so called Dirac points ${\bf Q}_i$, $i=1,\dots,6$. Near the Dirac points the electron dispersion is linear,  
\be  
E_{{\bf k}l}=(-1)^l  V|{\bf \tilde k}|, \ \ \tilde {\bf k}={\bf k}-  {\bf Q}_i, \ \ |\tilde {\bf k}|a\ll 1, \ \ V\simeq 10^6\ {\rm m/s},\label{spectrum}  
\ee  
and the behavior of electrons can be described by the Dirac equation with the Hamiltonian  
$ 
\hat H_D=V\sigma_\alpha\hat p_\alpha, 
$ 
where $\sigma_\alpha$ are Pauli matrices ($\alpha=x,y$). Only two of the six Dirac cones are physically inequivalent which is accounted for by the valley degeneracy factor $g_v=2$.

The linear (massless) dispersion of graphene electrons near the Fermi level leads to a number of interesting linear and non-linear electrodynamic phenomena, which are briefly discussed in this paper. Some of these effects are very promising for microwave and terahertz applications of graphene. 

\section{Linear electromagnetic response} 
\label{linresp} 

\subsection{Frequency-dependent conductivity} 

\begin{figure}  
\includegraphics[width=6.7cm]{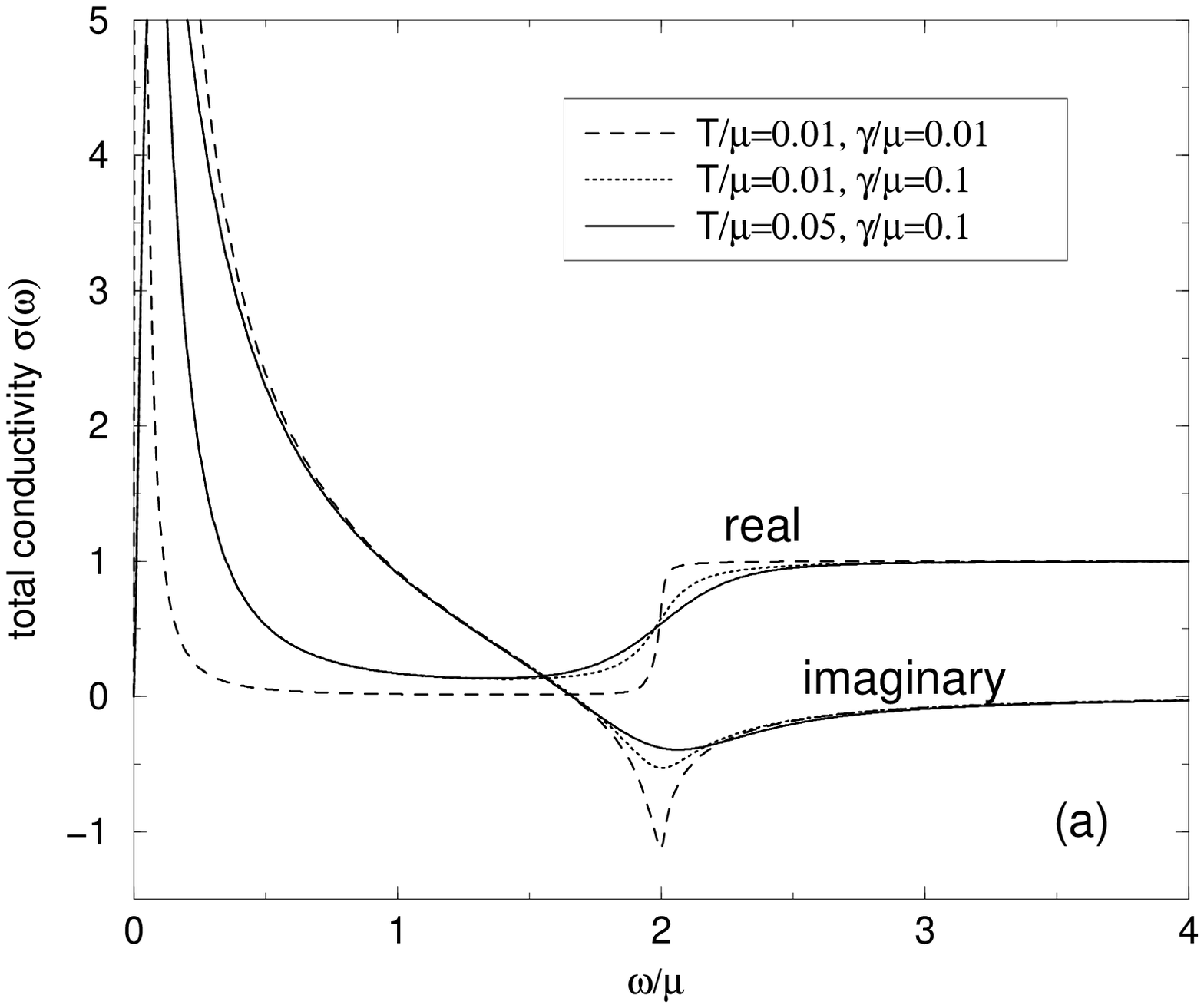}\hfill\includegraphics[width=6.7cm]{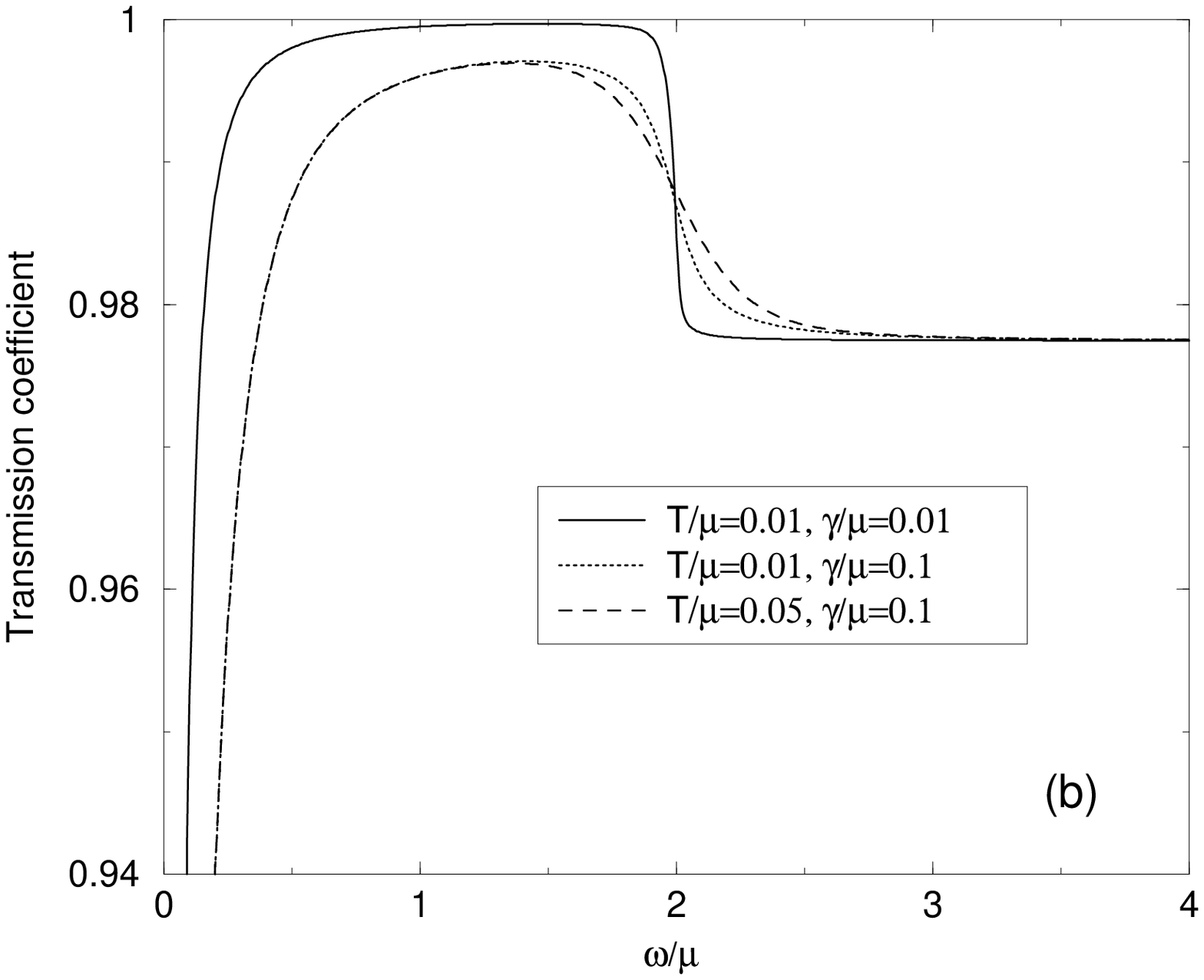}\\  
\caption{\label{cond} {\bf (a)} The real and imaginary parts of the conductivity $\sigma(\omega)$, in units $e^2/4\hbar$, and {\bf (b)} the transmission coefficient  of graphene as a function of $\hbar\omega/\mu$.
}  
\end{figure}

Using the standard self-consistent linear-response theory and starting from the Dirac Hamiltonian $\hat H_D$ one can calculate the frequency dependent conductivity of graphene $\sigma(\omega)=\sigma_{intra}(\omega)+\sigma_{inter}(\omega)$ \cite{Ando02,Falkovsky07a,Falkovsky07b,Gusynin06a,Peres06,Mikhailov07d}. It consists of two contributions. The {\em intra}-band conductivity  
\be 
\sigma_{intra}(\omega) 
= 
\frac {ie^2g_sg_vT}{2\pi\hbar^2 (\omega+i\gamma)} \ln \left(e^{\mu/2T}+e^{-\mu/2T}\right) 
\label{sigma-intra} 
\ee 
has the Drude form and is similar to the conductivity of conventional 2D electron systems. The {\em inter}-band conductivity has the form 
\ba 
\sigma_{inter}(\omega)&=& 
-\frac {i e^2 g_sg_v}{8\pi\hbar}\int_0^\infty  
\frac {\sinh x }{\cosh\mu/T+\cosh x} 
\times\frac {\hbar(\omega+i\gamma)/2T} 
{x^2-[\hbar(\omega+i\gamma)/2T]^2}dx ,
\label{sigma-inter} 
\ea 
where $T$, $\mu$ and $\gamma$ are the temperature, the chemical potential and the momentum scattering rate respectively, and $g_s=2$ is the spin degeneracy.  
Figure \ref{cond}a shows the total graphene conductivity $\sigma(\omega)$ as a function of $\omega/\mu$ at several typical values of $T/\mu$ and $\gamma/\mu$. At high frequencies, $\omega\gg \max\{T,\mu\}$, $\sigma(\omega)$ tends to a universal value $\sigma_{\omega\to\infty}\to e^2/4\hbar$ dependent only on the fundamental physical constants. The behaviour of $\sigma(\omega)$ shown in Figure \ref{cond}a has been  experimentally confirmed in Ref. \cite{Li08}.

\subsection{Transmission, reflection and absorption of light} 

The experimentally measurable transmission $T=|t|^2$, reflection $R=|r|^2$ and absorption $A=1-T-R$ spectra are determined by the conductivity $\sigma(\omega)$, since the transmission amplitude $t$ is $t=1+r=[1+2\pi\sigma(\omega)/c]^{-1}$. At high frequencies the transmission and absorption coefficients are universal, 
\be 
T=1-A=(1+\pi\alpha/2)^{-2}\approx 1-\pi\alpha, \ \omega\gg \max\{T,\mu\},\label{transm} 
\ee 
where $\alpha=e^2/\hbar c$. Figure \ref{cond}b illustrates the $T(\omega)$ dependence \cite{Falkovsky07b}. The universal transmission (\ref{transm}) has been  observed in \cite{Nair08}. 

\subsection{Plasmons and other electromagnetic excitations} 

The plasma waves in graphene have been studied in Refs. \cite{Hwang07,Wunsch06,Vafek06}. In the long-wavelength limit $q\ll k_F$ the 2D plasmons in graphene have a square-root dispersion ($k_F$ is the Fermi wave-vector)
\be 
\omega^2_p=g_sg_ve^2\mu q/2\hbar^2\propto \sqrt{n_s}q.\label{2dp} 
\ee 
The density ($n_s$) dependence of the 2D plasmon frequency in graphene $\omega_p\propto n_s^{1/4}$ differs from that of the conventional 2D plasmons ($\omega\propto n_s^{1/2}$). At larger wavevectors ($q\simeq k_F$, but still $qa\ll 1$) the 2D plasmons acquire an additional damping due to the inter-band absorption \cite{Wunsch06}.  

At even larger wavevectors ${\bf q}\simeq {\bf Q}_i$ ($q\simeq 1/a$) the 2D plasmon spectrum should be calculated taking into account the full energy spectrum of graphene electrons (\ref{sp}) and the local field effects \cite{Adler62}. Such calculations \cite{Mikhailov08c} reveal a new type of low-frequency plasmons -- the {\em inter-valley plasmons} -- with the linear dispersion $ \omega = S|{\bf q-Q}_i|$, where $i=1,\dots,6$ and $S\gtrsim V$. 

The 2D plasmons (\ref{2dp}) 
are the transverse magnetic (TM) modes. 
The transverse electric (TE) electromagnetic modes do not exist in conventional 2D electron systems. It was shown however \cite{Mikhailov07d} that in graphene the TE modes should exist at $\omega\lesssim 2\mu$, where the imaginary part of the dynamical conductivity $\sigma''(\omega)$ is negative, Figure \ref{cond}b. 

\section{Non-linear electromagnetic response} 
\label{nonlin} 

\subsection{Frequency multiplication} 

If a particle with the linear dispersion (\ref{spectrum}) is placed in the external electric field ${\bf E}(t)={\bf E}_0\cos\omega t$, its momentum ${\bf p}(t)$ will be proportional to $\sin\omega t$ and the velocity ${\bf v}(t)$ -- to ${\bf v}=\nabla_{\bf p} E({\bf p})=V {\bf p}/|{\bf p}|\sim \sgn(\sin\omega t)$. As the function  
\be 
\sgn(\sin\omega t)=\frac 4\pi\left(\sin\omega t +\frac 13\sin3\omega t+\frac 15\sin5\omega t+\dots\right) \label{sgnsin}
\ee  
contains all odd Fourier harmonics, irradiation of graphene by a wave with the frequency $\omega$ should lead (in contrast to the conventional electron systems) to the emission at higher harmonics at frequencies $m\omega$ with $m=1,3,5,\dots$ \cite{Mikhailov07e}.  

\begin{figure}  
\includegraphics[width=7cm]{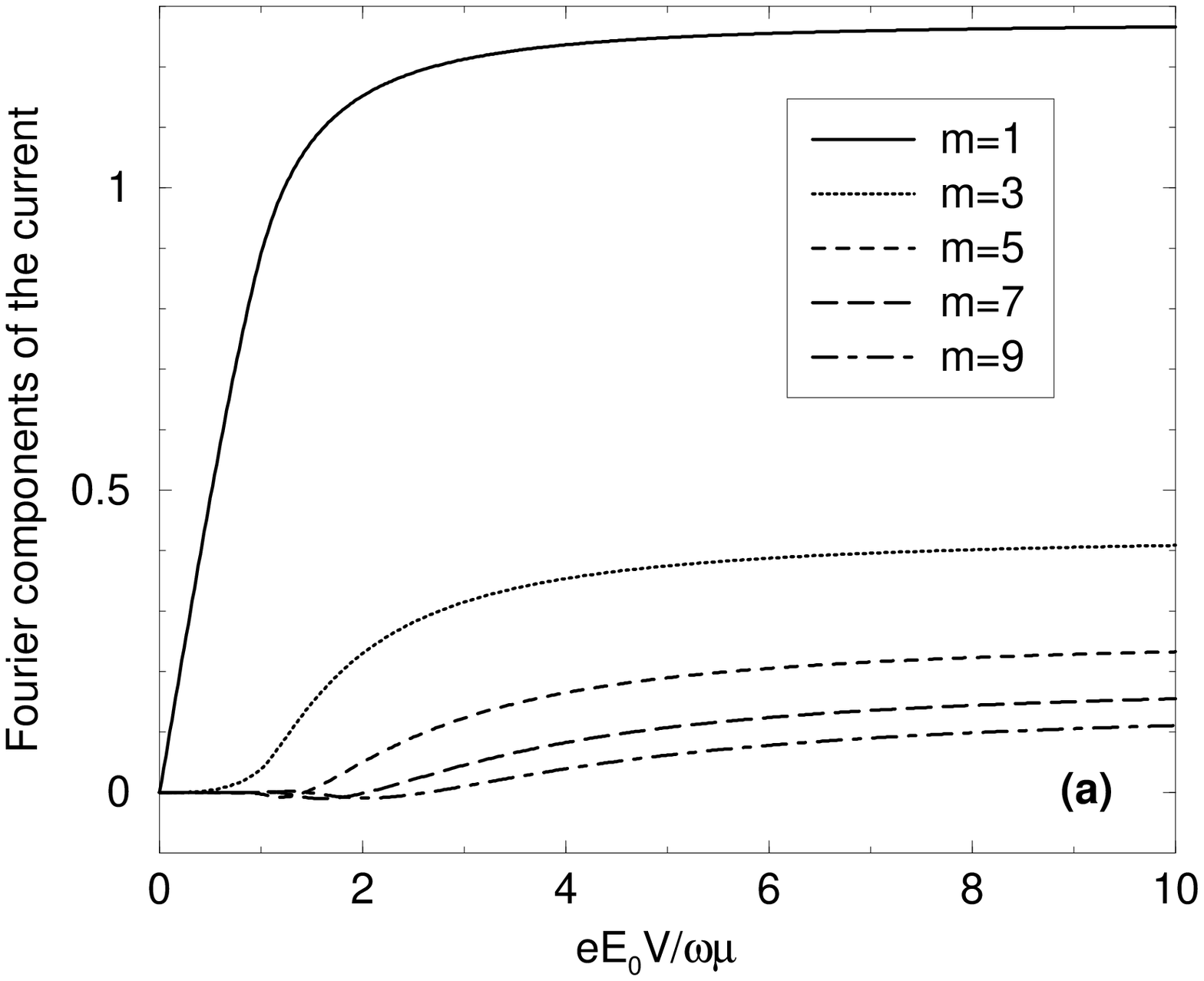}\hfill\includegraphics[width=7cm]{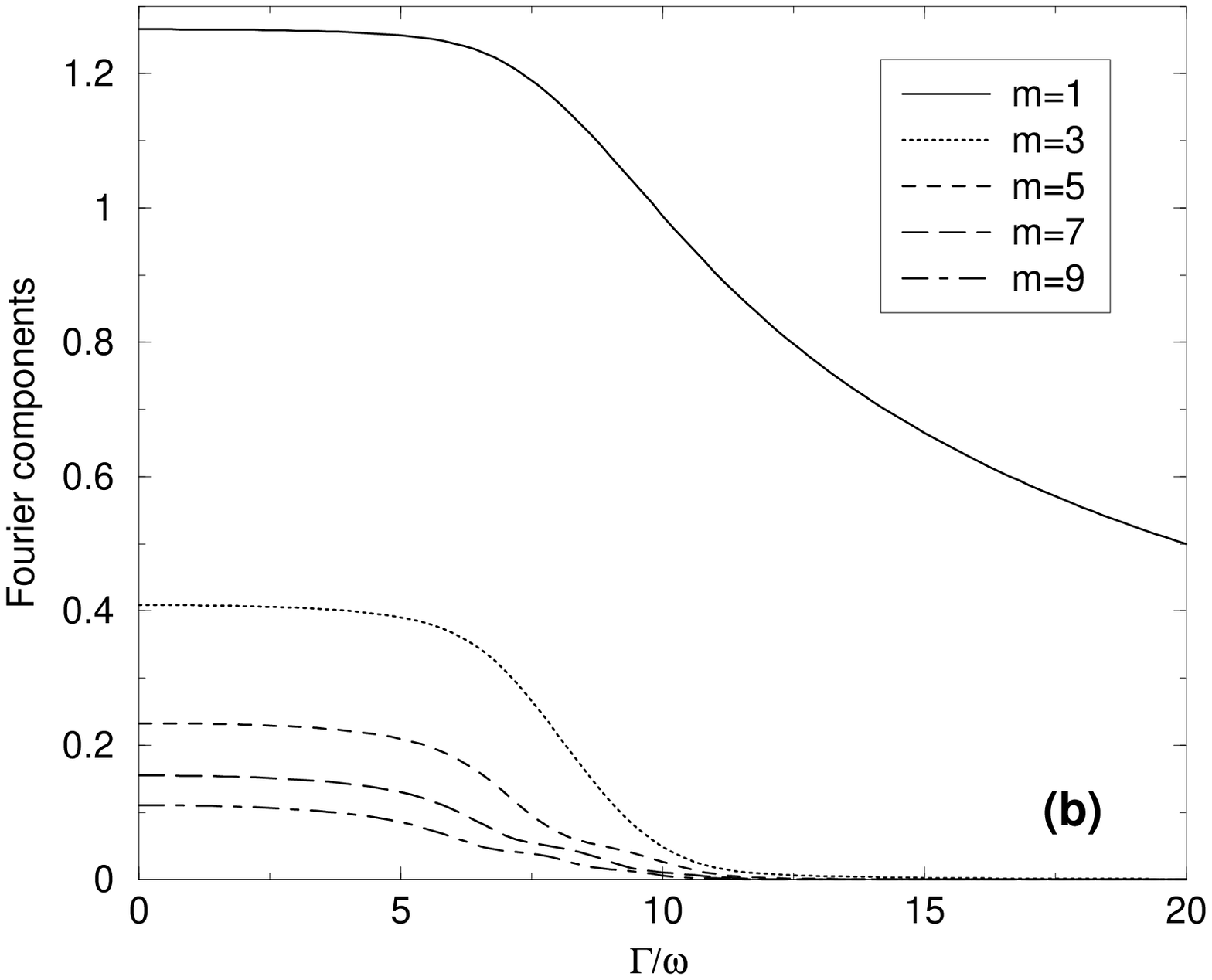}  
\caption{\label{mult} The Fourier components of the current $j/en_sV$ as a function of {\bf (a)} the field parameter ${\cal E}=eE_0V/\omega\mu$ at $\Gamma/\omega=0$ and {\bf (b)} of $\Gamma/\omega$ at ${\cal E}=10$. 
}  
\end{figure}

A more accurate theory \cite{Mikhailov08a} takes into account the distribution of electrons over quantum states in the energy bands and the self-consistent field effects. If $T\ll\mu$, the higher harmonics generation depends on two parameters ${\cal E}=eE_0V/\omega\mu$ and $\Gamma/\omega$, where $\Gamma=2\alpha \mu/\hbar$ is the (linear-response) radiative decay rate \cite{Mikhailov08a} (here $\alpha=e^2/\hbar c$). Our calculations show (Figure \ref{mult}) that the system efficiently generates higher harmonics if ${\cal E}\gtrsim \max\{1,\Gamma/\Omega\}$, i.e. at 
\be 
E_0\gtrsim\frac{\mu\Gamma}{eV}=\frac{2\pi n_seV}c\approx 300\frac{\rm V}{\rm cm}\times n_s(10^{11}/{\rm cm}^2).\label{estim33} 
\ee 
The estimate (\ref{estim33}) does not depend on the frequency of radiation (it is assumed however that the dimensions of the sample exceed the wavelength of radiation and that $\hbar\omega\lesssim 2\mu$). 

\subsection{Response to a pulse excitation}

Electromagnetic response of graphene to a strong pulse excitation ${\bf E}(t)= {\bf E}_{0}\tau_0\delta(t)$ also differs from that of conventional 2D electron systems (here ${\bf E}_{0}$ and $\tau_0$ are the amplitude and the duration of the pulse). The momentum relaxation in graphene after the strong pulse excitation is {\em linear} in time \cite{Mikhailov08a}, in contrast to the exponential relaxation in conventional systems. The characteristic response time in the non-linear regime is $\simeq eE_0\tau_0V/\mu\Gamma$. 

\subsection{Non-linear response in a magnetic field} 

To describe the influence of a magnetic field ${\bf B}=(0,0,B)$ on the non-linear electromagnetic response of graphene we solve the system of quasi-classical non-linear equations of motion for $N\gg 1$ Dirac quasi-particles
\be
\frac{d {\cal P}_j}{d\tau}=i{\cal B}\frac{{\cal P}_j}{|{\cal P}_j|}-{\cal E}\cos\tau-\frac\Gamma\omega {\cal J}, \ \ {\cal J}=\frac 1N
\sum_{k=1}^N\frac{{\cal P}_k}{|{\cal P}_k|}.\label{Bf}
\ee
Here $ {\cal P}_j=(p_x+ip_y)_j/p_F$ is the complex momentum of the $j$-th particle, normalized to the Fermi momentum $p_F$, ${\cal J}=(j_x+ij_y)/en_sV$ is the current, $\tau=\omega t$ and ${\cal B}=eBV^2/\omega\mu c$. The  term $\Gamma{\cal J}/\omega$ in (\ref{Bf}) results from the self-consistent-field effects and describes the radiative decay. 

\begin{figure}  
\includegraphics[width=7cm]{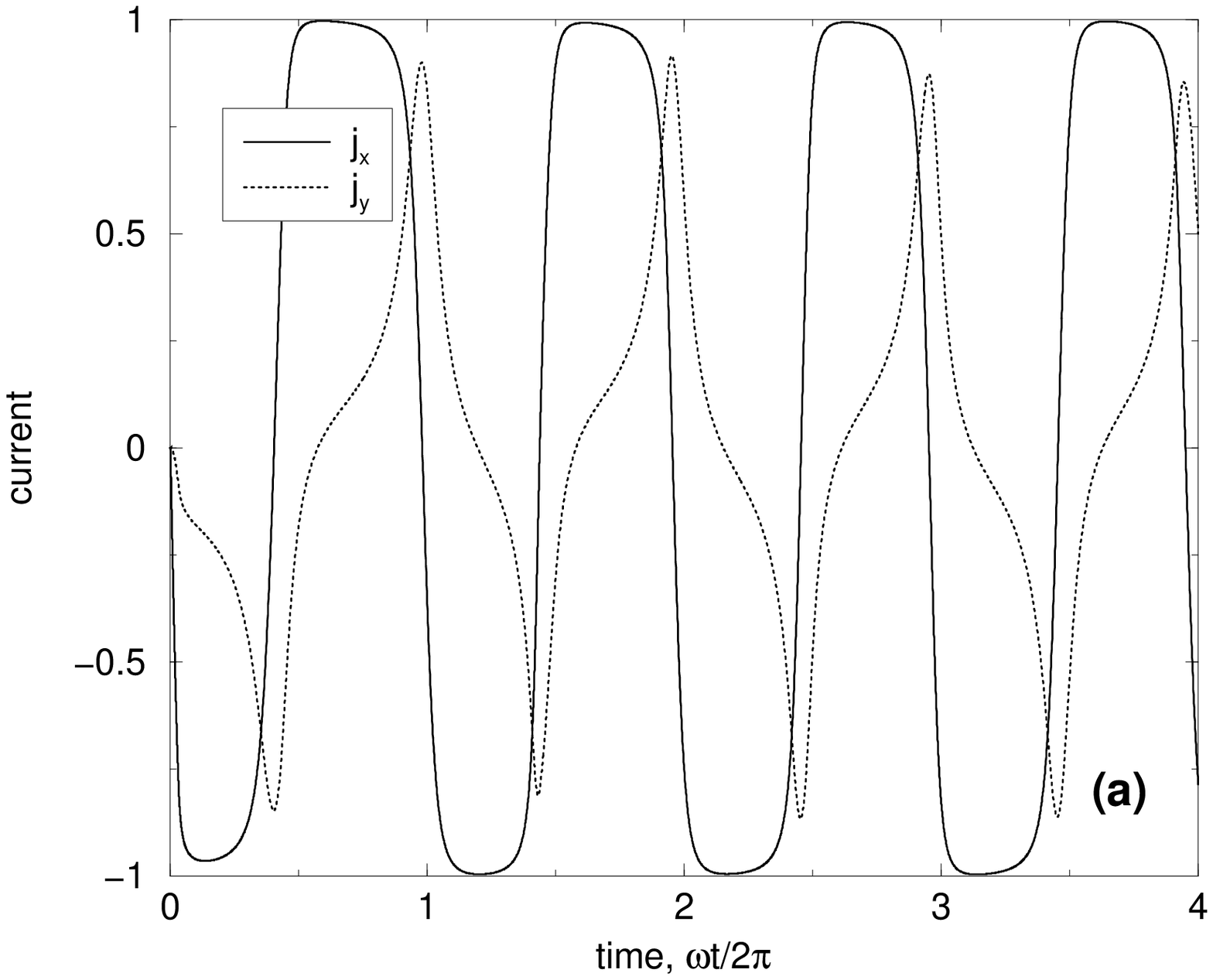}\hfill\includegraphics[width=7cm]{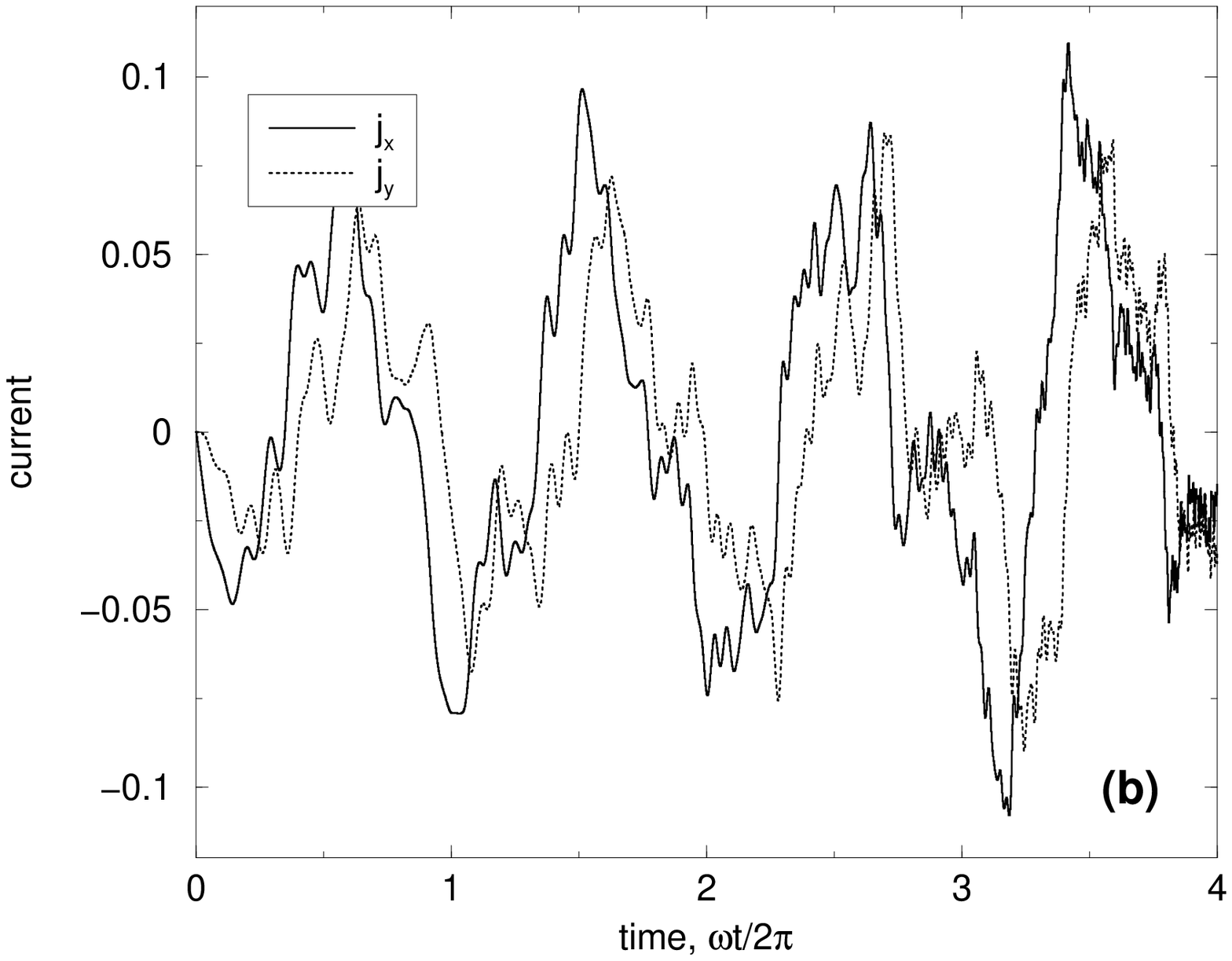}  
\caption{\label{Bfield} Electric current $j(t)/en_sV$ vs time at {\bf (a)} ${\cal E}=5$ and {\bf (b)} ${\cal E}=0.1$ and at ${\cal B}=1$, $\Gamma/\omega=1$. The $E$-field is polarized in the $x$-direction. 
}  
\end{figure}

Figure \ref{Bfield}a shows the time dependence of the ac electric current in the regime of the strong ac electric ${\cal E}\gg1$ and weak magnetic fields ${\cal B}\ll{\cal E}$. Apart from the  current $j_x(t)$ with the time dependence close to (\ref{sgnsin}) the system generates the Hall current $j_y(t)$ with the higher frequency harmonics. In the regime of weak electric fields ${\cal E}\ll{\cal B}$ the frequency transformation effects become even more complicated, Figure \ref{Bfield}b. The regular particle dynamics becomes chaotic and the system generates both lower and a lot of higher harmonics. This is a consequence  of the singular motion of graphene electrons in the vicinity of the Dirac point where the cyclotron frequency $\omega_c=eBV/pc$ of individual particles diverges at $p\to 0$. The transition  to the chaotic dynamics should be observed at $ E_0\lesssim VB/c$. 

\section{Summary}\label{summ} 

Due to the linear energy dispersion (\ref{spectrum}), graphene is a strongly non-linear material from the viewpoint of its electrodynamic properties. The frequency multiplication effect falls down very slowly with the harmonics index $m$ and should be seen in moderate electric fields. In weak magnetic fields it should be possible to observe the transition from the regular to chaotic particle dynamics dependent on the amplitudes of the $B$ and $E$ fields. The predicted non-linear electrodynamic phenomena open up new interesting opportunities for studying the fundamental physics of Dirac quasi-particles as well as for  building innovative devices for microwave and terahertz optoelectronics. 

I thank Igor Goychuk and Timur Tudorovskiy for useful discussions. The work was supported by the Swedish Research Council and INTAS.








\end{document}